\documentclass[conference,compsoc]{IEEEtran}

\usepackage{listings}

\ifCLASSOPTIONcompsoc
  \usepackage[nocompress]{cite}
\else
  \usepackage{cite}
\fi

\ifCLASSINFOpdf
\else
\fi
\usepackage{array}

\usepackage{hyperref}

\begin{document}
\bstctlcite{IEEEexample:BSTcontrol}

%
\title{Design Patterns which Facilitate Message Digest Collision Attacks on Blockchains}

\author{\IEEEauthorblockN{Peter Robinson}
\IEEEauthorblockA{School of Information Technology and Electrical Engineering, University of Queensland, Brisbane, Australia\\
Email: peter.robinson@uqconnect.edu.au}}


\maketitle

\begin{abstract}
Message digest algorithms are one of the underlying building blocks of blockchain platforms such as Ethereum. This paper analyses situations in which the message digest collision resistance property can be exploited by attackers. Two mitigations for possible attacks are described: longer message digest sizes make attacks more difficult; and, including timeliness properties limits the amount of time an attacker has to determine a hash collision.
\end{abstract}


%
\IEEEpeerreviewmaketitle

\section{Introduction}
Message digest algorithms are hard coded into the Ethereum platform design, being seen as unchanging, un-breakable building blocks. However, the effective security provided by message digest algorithms weakens over time due to improved cryptanalysis, increases in computational power, and the emergence of quantum computers. Though attacks may be many years away from being practical, it is important to understand how they can be mitigated so that design patterns which facilitate these attacks are not incorporated into future blockchain designs.

The \textit{collision resistance} message digest property states that it is difficult to determine $x_1$ and $x_2$ such that $h(x_1) = h(x_2)$ and $x_1 \neq x_2$. This property is significantly weaker than the \textit{preimage resistance} property (given $y$, it is difficult to determine $x$ such that $y = h(x)$) and the \textit{second preimage resistance} property (given $y$ and $x_1$, it is difficult to determine $x_2$ such that $y = h(x_1) = h(x_2)$ and $x_1 \neq x_2$). As such, the \textit{collision resistance} property should be prioritized for analysis.

\section{Exploitation Scenarios}
The scenarios in which the  message digest collision resistance property can be exploited are when:
\begin{itemize}  
\item The attacker chooses the value to be digested.
\item The attacker is able to trick an entity into doing something based the message digest and value.
\item There is no time-sensitive aspect to the value.
\end{itemize}
Time-sensitivity is important as the first message digest collisions for an algorithm have historically taken a significant period of time.

\section{Attacks}
Two example attack scenarios which match the \textit{Exploitation Scenarios} are now described. Based on the proposal for EIP86 \cite{buterin2016c} as at May 2017, when a Transaction Verification Contract is deployed the address is calculated using the equation: $Keccak256(rlp.encode([creator + nonce + initcode])) \% 2^{160}$, where the $creator$ is the account that created the contract the $nonce$ is a value which should be sequential, and $initcode$ is the initialization code for the contract. An attacker who had broken the message digest collision resistance property could manipulate the $nonce$ value or the contents of the contract such that there exists two contracts which would be deployed to the same address. The attacker could show one variant of the contract to a user and persuade them to send some Ether to the address. The attacker could then deploy the nefarious contract and spend the Ether.

Ethereum's Proof of Stake proposal \cite{ethereum2017a} needs an agreed \textit{random number} to use to choose which miner will create the next block. One proposal is that a random value be generated by each miner, the miners submit message digest commitment values prior to exposing their random values to prevent cheating. The random values are then combined using XOR to produce the agreed \textit{random number}. An attacker who had broken the collision resistance property could determine two random values which hash to produce the same commitment value. They could submit their commitment value and then, after seeing the random values from the other miners, choose which of their random values to deliver, thus influencing the resulting agreed \textit{random number}.

\section{Recommended Attack Mitigations}
The recommended attack mitigations are:

\begin{itemize}
\item The message digest size used for account numbers should be increased. In an email from V. Buterin and M. Swende from the Ethereum team on May 18, 2017, they tentatively agreed to adopt this recommendation of mine.
\item A temporal component should be incorporated into designs. The random value proposed for use in PoS should incorporate some value from a recent block.
\end{itemize}




\bibliographystyle{IEEEtran}
\bibliography{IEEEabrv,paper1}

%
%
%


\end{document}